\documentclass[12pt]{article}
\usepackage{mitpress}
\usepackage{amsmath}
\usepackage{graphicx}
\usepackage{amsfonts}
\usepackage{amssymb}

\newdimen\dummy
\dummy=\oddsidemargin
\begin{document}

\title{Information, Relative Entropy of Entanglement and Irreversibility}
\author{L. Henderson and V. Vedral\\Centre for Quantum Computation, Clarendon Laboratory, University of Oxford,\\Parks Road OX1 3PU}
\maketitle
\begin{abstract}
Previously proposed measures of entanglement, such as entanglement of
formation and assistance, are shown to be special cases of the relative
entropy of entanglement. The difference between these measures for an ensemble
of mixed states is shown to depend on the availability of classical
information about particular members of the ensemble. Based on this, relations
between relative entropy of entanglement and mutual information are derived.
\end{abstract}

In quantifying entanglement, a number of measures have been proposed. For
bipartite pure states, $\rho_{AB}$, the Von Neumann entropy of the reduced
density matrix of either subsystem, $\rho_{A}$ or $\rho_{B}$, has been found
to be a good and unique measure, \cite{Bennett96},\cite{Popescu97}. Relative
entropy of entanglement has been proposed as a measure which extends to mixed
states, \cite{Vedral97PRL}, \cite{Vedral98}. Loosely speaking, it quantifies
how `far' an entangled state is from the set of disentangled states.
Entanglement of mixed states has also been characterised by the `entanglement
of formation', \cite{Smolin96},\cite{Wootters98}, and by the `entanglement of
distillation', \cite{Smolin96}. Rather surprisingly, use of entanglement in
mixed states is not reversible in the sense that all the entanglement required
to construct a particular mixed state cannot be distilled out again, so the
entanglement of formation is greater than the entanglement of distillation,
\cite{Vedral98}. In this paper, we clarify the role of classical information
about the identity of particular members of an ensemble of mixed states, and
show that the loss of such information is responsible for the difference
between the entanglement of formation and the entanglement of
distillation.\ We provide a unifying frame-work for entanglement measures by
showing how previously proposed measures are special cases of the relative
entropy of entanglement. This gives a strong physical argument for using
quantum relative entropy as a unique way to understand entanglement in general.

Suppose that Alice and Bob share a state described by the density matrix
$\rho_{AB}$. The state $\rho_{AB}$ has an infinite number of different
decompositions $\varepsilon=\{\left|  \psi_{AB}^{i}\right\rangle \langle
\psi_{AB}^{i}|,p_{i}\}$, into pure states $\left|  \psi_{AB}^{i}\right\rangle
$, with probabilities $p_{i}$, \cite{Hughston93}. We denote the mixed state
$\rho_{AB}$ written in decomposition $\varepsilon$ by
\begin{equation}
\rho_{AB}^{\varepsilon}=\sum_{i}p_{i}\left|  \psi_{AB}^{i}\right\rangle
\langle\psi_{AB}^{i}|\label{eq:decomp}%
\end{equation}
The entanglement of formation is the average entanglement of the pure states,
minimised over all decompositions, \cite{Wootters98}:%
\begin{equation}
E_{F}(\rho_{AB})=\min_{\varepsilon}\sum_{i}p_{i}S(\rho_{B}^{i}%
)\label{eq:entform}%
\end{equation}
Here $\rho_{B}^{i}$ is the reduced density matrix for subsystem $A$ of the
pure state $\left|  \psi_{AB}^{i}\right\rangle \langle\psi_{AB}^{i}|$. The
physical importance of entanglement of formation lies in the fact that it is
possible to convert an ensemble of $\ m$ maximally entangled singlets into a
smaller number, $n$, of non-maximally entangled states, $\rho_{AB}^{\otimes
n}$, using only local operations and classical communication, \cite{Smolin96},
and entanglement of formation is the asymptotic conversion ratio, $\frac{m}%
{n}$, in the limit of infinitely many copies.

The `entanglement of distillation', $E_{D}(\rho_{AB})$, is the number of
maximally entangled singlets per copy of $\rho_{AB\text{ }}$which can be
distilled from an asymptotically large ensemble of copies of $\rho_{AB}$ by a
purification procedure involving only local operations and classical
communication, \cite{Smolin96}. For a mixed state, it is lower than or equal
to the entanglement of formation, \cite{Vedral98}.

Relative entropy of entanglement of the mixed state is defined as
\[
E_{RE}(\rho_{AB})=\min_{\sigma_{AB}\in D}S(\rho_{AB}||\sigma_{AB})
\]
where $S(\rho||\sigma)=Tr(\rho\log\rho-\rho\log\sigma)$ is the quantum
relative entropy, \cite{Vedral98}. The minimum is taken over $D$, the set of
completely disentangled or `separable' states. A state is separable if it can
be written as a convex combination of product\nolinebreak
\ states\nolinebreak \ $\sigma=\sum_{i}p_{i}\sigma_{A}^{i}\otimes\sigma
_{B}^{i}$, with $\sum_{i}p_{i}=1$. The relative entropy of entanglement
provides an upper bound for the distillable entanglement, \cite{Vedral98}. The
known relationships between the different measures of entanglement for mixed
states are therefore $E_{D}(\rho_{AB})\leq E_{RE}(\rho_{AB})\leq E_{F}%
(\rho_{AB})$. Equality holds for pure states, where all the measures reduce to
the Von Neumann entropy, $S(\rho_{A})=S(\rho_{B})$. We will give a
straight-forward argument for the second inequality later in the paper.

Protocols for formation and distillation of pure and mixed entangled states
have been introduced \cite{Bennett96}, \cite{Smolin96}. We first briefly
review the pure state procedures, and then go on to discuss the role of
classical information and relative entropy of entanglement in the mixed state
case. 

The basis of formation is that Alice and Bob would like to create an ensemble
of $n$ copies of the non-maximally entangled state, $\rho_{AB}$, using only
local operations, classical communication, and a number of maximally entangled
pairs. It is customary to consider the process of formation which consumes the
least entanglement, since the only `cost' in communication is due to the use
of entanglement resources, or sending information down a quantum channel, and
classical communication costs nothing. Distillation is the reverse process,
where Alice and Bob share an ensemble of $n$ copies of the non-maximally
entangled state, $\rho_{AB}$, and would like to extract the largest number of
maximally entangled pairs using only local operations and classical communications.

Formation of an ensemble of $n$ non-maximally entangled \textit{pure} states,
$\rho_{AB}=\left|  \psi_{AB}\right\rangle \langle\psi_{AB}|$ is achieved by
the following protocol. Alice first prepares the states she would like to
share with Bob locally. She then uses Schumacher compression,
\cite{Schumacher95}, to compress these states into $nS(\rho_{B})$ states. One
particle of each pair is then teleported to Bob using $nS(\rho_{B})$ maximally
entangled pairs. Bob decompresses the states he receives and so ends up
sharing $n$ copies of $\rho_{AB}$ with Alice. The entanglement of formation is
therefore $E_{F}(\rho_{AB})=S(\rho_{B})$. For pure states, this process
requires no classical communication in the asymptotic limit, \cite{Lo99}. The
reverse process of distillation is accomplished using the Schmidt projection
method, \cite{Bennett96}, which allows $nS(\rho_{B})$ maximally entangled
pairs to be distilled in the limit as $n$ becomes very large. Again, this
process involves no classical communication between the separated parties.
Therefore pure states are fully inter-convertible in the asymptotic limit.

The situation for mixed states is more complex. When any mixed state, denoted
by Eq.(\ref{eq:decomp}), is created, it is necessarily part of an extended
system whose state is pure. We will consider the pure states $|\psi_{AB}%
^{i}\rangle$ in the mixture to be correlated to orthogonal states
$|m_{i}\rangle$ of a memory $M$. The extended system is in the pure state%
\[
\left|  \psi_{MAB}\right\rangle =\sum_{i}\sqrt{p_{i}}|m_{i}\rangle|\psi
_{AB}^{i}\rangle
\]
If we have no access to the memory system, we trace over it to obtain the
mixed state in Eq.(\ref{eq:decomp}). We will see that the amount of
entanglement involved in the different entanglement manipulations of mixed
states depends on the accessibility of the information in the memory at
different stages.

Note that a unitary operation on $\left|  \psi_{MAB}\right\rangle $ will
convert it into another pure state $\left|  \phi_{MAB}\right\rangle $ with the
same entanglement, \cite{Lindblad75},
\begin{equation}
\left|  \phi_{MAB}\right\rangle =\sum_{j}\sqrt{q_{j}}|n_{j}\rangle|\phi
_{AB}^{j}\rangle\label{eq:pure2}%
\end{equation}
Tracing over the memory in this case gives another decomposition,\allowbreak
\linebreak \ $\zeta=\{\left|  \phi_{AB}^{j}\right\rangle \langle\phi_{AB}%
^{j}|,q_{j}\}$, of $\rho_{AB}$ into pure states%
\begin{equation}
\rho_{AB}^{\zeta}=\sum_{j}q_{j}\left|  \phi_{AB}^{j}\right\rangle \langle
\phi_{AB}^{j}|\label{eq:mixed2}%
\end{equation}
The reduction of the pure state, (\ref{eq:pure2}), to the mixed state,
(\ref{eq:mixed2}), may be regarded as due to a projection-valued measurement
on the memory with operators $\{E_{j}=\left|  n_{j}\right\rangle \langle
n_{j}|\}$.

Consider first the protocol of formation by means of which Alice and Bob come
to share an ensemble of $n$ mixed state $\rho_{AB}$. Alice first creates the
mixed states locally by preparing a collection of $n$ states in a particular
decomposition, $\varepsilon=\{\left|  \psi_{AB}^{i}\right\rangle \langle
\psi_{AB}^{i}|,p_{i}\}$ by making $np_{i}$ copies of each pure state
$|\psi_{AB}^{i}\rangle$. At the same time a memory system entangled to the
pure states is generated, which keeps track of the identity of each member of
the ensemble. Note that as long as we consider Alice's entire environment, the
state of subsystems $A$ and $B$ together with the memory may always be taken
to be pure. Later, we will consider the situation in which Alice's memory is
decohered. There are then three ways for her to share these states with Bob.
First of all, she may simply compress subsystem $B$ to $nS(\rho_{B})$ states,
and teleport these to Bob using $nS(\rho_{B})$ maximally entangled pairs. The
choice of which subsystem to teleport is made so as to minimise the amount of
entanglement required, so that $S(\rho_{B})\leq S(\rho_{A})$. The
teleportation in this case would require no classical communication in the
asymptotic limit, just as for pure states, \cite{Lo99}. The state of the whole
system which is created by this process is an ensemble of pure states $\left|
\psi_{MAB}\right\rangle $, where subsystems $M$ and $A$ are on Alice's side
and subsystem $B$ is on Bob's side. In terms of entanglement resources,
however, this process is not the most efficient way for Alice to send the
states to Bob. She may do it more efficiently by using the memory system of
$\left|  \psi_{MAB}\right\rangle $ to identify blocks of $np_{i\text{ }}%
$members in each pure state $\left|  \psi_{AB}^{i}\right\rangle $, and
applying compression to each block to give $np_{i}S(\rho_{B}^{i})$ states.
Then the total number of maximally entangled pairs required to teleport these
states to Bob is $n\sum_{i}p_{i}S(\rho_{B}^{i})$, which is clearly less than
$nS(\rho_{B})$, by concavity of the entropy. The amount of entanglement
required clearly depends on the decomposition of the mixed state $\rho_{AB}$.
However, in order to decompress these states, Bob must also be able to
identify which members of the ensemble are in which state. Therefore Alice
must also send him the memory system. She now has two options. She may either
teleport the memory to Bob, which would use more entanglement resources. Or
she may communicate the information in the memory classically, with no further
use of entanglement. When Alice uses the minimum entanglement decomposition,
$\varepsilon=\{\left|  \psi_{AB}^{i}\right\rangle \langle\psi_{AB}^{i}%
|,p_{i}\}$, this process, originally introduced by Bennett \textit{et al.},
\cite{Smolin96}, makes the most efficient use of entanglement, consuming only
the entanglement of formation of the mixed state, $E_{F}(\rho_{AB})=\sum
_{i}p_{i}S(\rho_{B}^{i})$. We may think of the classical communication between
Alice and Bob in one of two equivalent ways. Alice may either measure the
memory locally to decohere it, and then send the result to Bob classically, or
she may send the memory through a completely decohering quantum channel. In
this case, the interaction with the channel is given by%
\begin{align*}
\left|  \psi_{MAB}\right\rangle \left|  c\right\rangle  &  =\sum_{i}%
\sqrt{p_{i}}|m_{i}\rangle|\psi_{AB}^{i}\rangle\left|  c\right\rangle \\
&  \longrightarrow\sum_{i}\sqrt{p_{i}}|\psi_{AB}^{i}\rangle\left|
m_{i}\right\rangle \left|  c_{i}\right\rangle
\end{align*}
where $\left|  c\right\rangle $ is the initial state of the channel and
$\{\left|  c_{i}\right\rangle \}$ are orthogonal channel states. Since Alice
and Bob have no access to the channel, the state of the whole system which is
created by this process is the mixed state
\begin{equation}
\rho_{ABM}^{\varepsilon}=\sum_{i}p_{i}|\psi_{AB}^{i}\rangle\langle\psi
_{AB}^{i}|\otimes|m_{i}\rangle\langle m_{i}|\label{eq:mixed}%
\end{equation}
where Bob is classically correlated to the $AB$ subsystem. Bob is then able to
decompress his states using the memory to identify members of the ensemble.

Once the collection of $n$ pairs is shared between Alice and Bob, it is
converted into an ensemble of $n$ mixed states $\rho_{AB}$ by destroying
access to the memory which contains the information about the state of any
particular member of the ensemble. It is the loss of this information which is
responsible for the fact that entanglement of distillation is lower than than
entanglement of formation\footnote{The relation between classical information
and distillable entanglement was previously discussed by Eisert \textit{et
al.}, \cite{Eisert99}, in a different context.}. Distillation is not carried
out by people like Alice and Bob who have access to the memory, but by people
who have just received the ensemble of $n$ mixed states $\rho_{AB}$ with no
further information. If Alice and Bob were to carry out the distillation, they
could obtain as much entanglement from the ensemble as was required to form
it. In the case where Alice and Bob share an ensemble of\ the pure state
$\left|  \psi_{MAB}\right\rangle $, they would simply apply the Schmidt
projection method, \cite{Bennett96}. The relative entropy of entanglement
gives the upper bound to distillable entanglement,
\begin{equation}
E_{RE}(\left|  \psi_{(MA):B}\right\rangle \langle\psi_{(MA):B}|)=S(\rho
_{B})\label{eq:repure}%
\end{equation}
which is the same as the amount of entanglement required to create the
ensemble of pure states, as described above. Here $MA$ and $B$ are spatially
separated subsystems on which joint operations may not be performed. In our
notation, we use a colon to separate the local subsystems. 

On the other hand, if Alice used the least entanglement for producing an
ensemble of the mixed state $\rho_{AB}$, together with classical
communication, the state of the whole system is an ensemble of the mixed state
$\rho_{ABM}^{\varepsilon}$, and the process is still reversible. Because of
the classical correlation to the states $\left|  \psi_{AB}^{i}\right\rangle $,
Alice and Bob may identify blocks of members in each pure state $\left|
\psi_{AB}^{i}\right\rangle $, and apply the Schmidt projection method to them,
giving $np_{i}S(\rho_{B}^{i})$ maximally entangled pairs, and hence a total
entanglement of distillation of $\sum_{i}p_{i}S(\rho_{B}^{i})$. The relative
entropy of entanglement again quantifies the amount of distillable
entanglement from the state $\rho_{ABM}^{\varepsilon}$ and is given by%

\[
E_{RE}(\rho_{A:(BM)}^{\varepsilon})=\min_{\sigma_{ABM}\in D}S(\rho
_{ABM}^{\varepsilon}||\sigma_{ABM})
\]
The disentangled state which minimises the relative entropy is \linebreak
$\sigma_{ABM}=\sum_{i}p_{i}\sigma_{AB}^{i}\otimes|m_{i}\rangle\langle m_{i}|$,
\noindent where $\sigma_{AB}^{i}$ is obtained from $|\psi_{AB}^{i}%
\rangle\langle\psi_{AB}^{i}|$ by deleting the off-diagonal elements in the
Schmidt basis. This is the minimum because the state $\rho_{MAB}$ is a mixture
of the orthogonal states $\left|  m_{i}\right\rangle |\psi_{AB}^{i}\rangle$,
and for a pure state $|\psi_{AB}^{i}\rangle$, the disentangled state which
minimises the relative entropy is $\sigma_{AB}^{i}$. The minimum relative
entropy of the extended system is then%
\begin{equation}
S(\rho_{ABM}^{\varepsilon}||\sigma_{ABM})=\sum_{i}p_{i}S(\rho_{B}%
^{i})\nonumber
\end{equation}
This relative entropy, $E_{RE}(\rho_{A:(BM)}^{\varepsilon})$, has previously
been called `entanglement of projection', \cite{Garisto98}, because the
measurement on the memory projects the pure state of the full system into a
particular decomposition. The minimum of $E_{RE}(\rho_{A:(BM)}^{\varepsilon})$
over all decompositions is equal to the entanglement of formation of
$\rho_{AB}$. However, Alice and Bob may choose to create the state $\rho_{AB}$
by using a decomposition with higher entanglement than the entanglement of
formation. The maximum of $E_{RE}(\rho_{A:(BM)}^{\varepsilon})$ over all
possible decompositions is called the `entanglement of assistance' of
$\rho_{AB}$, \cite{Divincenzo98}. Because $E_{RE}(\rho_{A:(BM)}^{\varepsilon
})$ is a relative entropy, it is invariant under local operations and
non-increasing under general operations, properties which are conditions for a
good measure of entanglement, \cite{Vedral98}. However, unlike $E_{RE}%
(\rho_{AB})$ and $E_{F}(\rho_{AB})$, it is not zero for completely
disentangled states. In this sense, the relative entropy of entanglement,
$E_{RE}(\rho_{A:(BM)}^{\varepsilon})$, defines a class of entanglement
measures interpolating between the entanglement of formation and entanglement
of assistance. Note that an upper bound for the entanglement of assistance,
$E_{A}$, can be shown using concavity, \cite{Divincenzo98}, to be $E_{A}%
(\rho_{AB})\leq\min[S(\rho_{A}),S(\rho_{B})]$. This bound can also be shown
from the fact that the distillable entanglement from any decomposition,
$E_{RE}(\rho_{A:(BM)}^{\varepsilon})\leq E_{A}(\rho_{AB})$ cannot be greater
than the entanglement of the original pure state.

We may also derive relative entropy measures that interpolate between the
relative entropy of entanglement and the entanglement of formation by
considering non-orthogonal measurements on the memory. First of all, the fact
that the entanglement of formation is in general greater than the upper bound
for entanglement of distillation, emerges as a property of the relative
entropy, namely that it cannot increase under the local operation of tracing
one subsystem, \cite{Lindblad75},%
\[
E_{F}(\rho_{AB})=\min_{\sigma_{ABM}\in D}S(\rho_{ABM}||\sigma_{ABM})\geq
\min_{\sigma_{AB}\in D}S(\rho_{AB}||\sigma_{AB})
\]
In general, the loss of the information in the memory may be regarded as a
result of an imperfect classical channel. This is equivalent to Alice making a
non-orthogonal measurement on the memory, and sending the result to Bob. In
the most general case, $\{E_{i}=A_{i}A_{i}^{+}\}$ is a POVM performed on the
memory. The decomposition corresponding to this measurement is composed of
mixed states, $\xi=\{q_{i},Tr_{M}(A_{i}\rho_{MAB}A_{i}^{+})\}$, where
$q_{i}=Tr(A_{i}\rho_{MAB}A_{i}^{+})$. The relative entropy of entanglement of
the state $\rho_{MAB}^{\xi}$, when $\xi$ is a decomposition of $\rho_{AB}$
resulting from a non-orthogonal measurement on $M$, defines a class of
entanglement measures interpolating between the relative entropy of
entanglement and the entanglement of formation of the state $\rho_{AB}$. In
the extreme case where the measurement gives no information about the state
$\rho_{AB}$, $E_{RE}(\rho_{A:(BM)}^{\varepsilon})$ becomes the relative
entropy of entanglement of the state $\rho_{AB}$ itself. In between, the
measurement gives partial information. We note that instead of an imperfect
measurement, we may regard the memory itself as imperfect, in the sense that
the memory states are non-orthogonal, $\langle m_{i}\left|  m_{j}\right\rangle
\neq0$ for $i\neq j$.

Now we will relate the loss of entanglement to the loss of information in the
memory. As we have argued so far, there are two stages at which distillable
entanglement is lost. The first is in the conversion of the pure state
$\left|  \psi_{MAB}\right\rangle $ into a mixed state $\rho_{ABM}$. This
happens because Alice uses a \textit{classical} channel to communicate the
memory to Bob. The second is due to the loss of the memory, $M$, taking the
state $\rho_{ABM}$ to $\rho_{AB}$. The amount of information lost may be
quantified by the difference in mutual information between the respective
states. Mutual information is a measure of correlations between the memory $M$
and the system $AB$, giving the amount of information about $AB$ which may be
obtained from a measurement on $M$. The quantum mutual information between $M$
and $AB$ is defined as
\[
I_{Q}(\rho_{M:(AB)})=S(\rho_{M})+S(\rho_{AB})-S(\rho_{MAB})\linebreak
\]
The quantum mutual information of the pure state $\left|  \psi_{MAB}%
\right\rangle $ is\linebreak \ $I_{Q}(\left|  \psi_{M:(AB)}\right\rangle
\langle\psi_{M:(AB)}|)=2S(\rho_{AB})$, and of the mixed state in
Eq.(\ref{eq:mixed}) is $I_{Q}(\rho_{M:(AB)})=S(\rho_{AB})$. Therefore the
mutual information loss in the first stage is $\Delta I_{Q}=S(\rho_{AB})$.
There is a corresponding reduction in the relative entropy of entanglement,
from the entanglement of the original pure state, $E_{RE}(\left|
\psi_{(MA):B}\right\rangle \langle\psi_{(MA):B}|)$, to the entanglement of the
mixed state $E_{RE}(\rho_{A:(BM)}^{\varepsilon})$ for all decompositions
$\varepsilon$ arising as the result of an orthogonal measurement on the
memory. We now show that when the mutual information loss is added to the
relative entropy of entanglement of the mixed state $E_{RE}(\rho
_{A:(BM)}^{\varepsilon})$, the result is greater than the relative entropy of
entanglement of the original pure state, $E_{RE}(\left|  \psi_{(MA):B}%
\right\rangle \langle\psi_{(MA):B}|)$. \ We show the result for the strongest
case, which occurs when $E_{RE}(\rho_{A:(BM)}^{\varepsilon})=E_{F}(\rho_{AB}%
)$:
\begin{equation}
E_{RE}(\left|  \psi_{(MA):B}\right\rangle \langle\psi_{(MA):B}|)\leq
E_{F}(\rho_{AB})+S(\rho_{AB})\label{eq:ptom}%
\end{equation}
The proof goes as follows. Let $\varepsilon=\{\left|  \psi_{AB}^{i}%
\right\rangle \langle\psi_{AB}^{i}|,p_{i}\}$ be the minimal entanglement
decomposition giving rise to the entanglement of formation, see
Eq.(\ref{eq:entform}). Then,
\begin{align*}
S(\rho_{AB}) &  =\sum_{i}p_{i}S(\left|  \psi_{AB}^{i}\right\rangle \langle
\psi_{AB}^{i}|||\rho_{AB})\\
&  \geq\sum_{i}p_{i}S(\rho_{B}^{i}||\rho_{B})\\
&  =S(\rho_{B})-\sum_{i}p_{i}S(\rho_{B}^{i})
\end{align*}
The inequality results from the fact that the relative entropy does not
increase under the local operation of tracing subsystem $A$, \cite{Lindblad75}%
. Using Eq.(\ref{eq:repure}), and the fact that for this decomposition,
$E_{F}(\rho_{AB})=$ $\sum_{i}p_{i}S(\rho_{B}^{i})$, gives inequality
(\ref{eq:ptom}).

A similar result may be proved for the second loss, due to loss of the memory.
After this, the mutual information between the memory and $AB$ of the state
$\rho_{AB}$ is zero. Therefore the mutual information lost in losing the
memory is again $\Delta I_{Q}=S(\rho_{AB})$. The relative entropy of
entanglement is reduced from $E_{RE}(\rho_{A:(BM)}^{\varepsilon})$, for any
decomposition $\varepsilon$ resulting from an orthogonal measurement on the
memory, to $E_{RE}(\rho_{AB})$, the relative entropy of entanglement of the
state $\rho_{AB}$ with no memory. We show that when the mutual information
loss is added to $E_{RE}(\rho_{AB})$, the result is greater than $E_{RE}%
(\rho_{A:(BM)}^{\varepsilon})$. In this case, the result is strongest for
$E_{RE}(\rho_{A:(BM)}^{\varepsilon})=E_{A}(\rho_{AB})$:%
\begin{equation}
E_{A}(\rho_{AB})\leq E_{RE}(\rho_{AB})+S(\rho_{AB}) \label{eq:upper}%
\end{equation}
Let $\zeta=\{\left|  \phi_{AB}^{i}\right\rangle \langle\phi_{AB}^{i}|,q_{i}\}$
be the maximal entanglement decomposition giving rise to the entanglement of
assistance. Then
\begin{align*}
E_{RE}(\rho_{AB})+S(\rho_{AB})-E_{RE}(\rho_{A:(BM)}^{\zeta})=  & \\
-\min_{\sigma_{AB}\in D}Tr(\rho_{AB}\ln\sigma_{AB})-\sum_{i}q_{i}S(\rho
_{B}^{i})  &  =\\
\qquad\quad\sum_{i}q_{i}(-\min_{\sigma_{AB}\in D}\langle\phi_{AB}^{i}%
|\ln\sigma_{AB}\left|  \phi_{AB}^{i}\right\rangle -S(\rho_{B}^{i}))  &  \geq\\
\sum_{i}q_{i}(\min_{\sigma_{AB}^{i}\in D}S(\left|  \phi_{AB}^{i}\right\rangle
\langle\phi_{AB}^{i}|||\sigma_{AB}^{i})-S(\rho_{B}^{i}))  &  =0
\end{align*}
The inequality holds because $\sigma_{AB}$ is the disentangled state which
minimises the relative entropy of the state $\rho_{AB}$, but may not minimise
the relative entropy for each of the component pure states, $\left|  \phi
_{AB}^{i}\right\rangle $. Notice that if $\rho_{AB}$ is a pure state, then
$S(\rho_{AB})=0$, and equality holds.

Inequalities (\ref{eq:ptom}) and (\ref{eq:upper}) provide lower bounds for
$E_{F}(\rho_{AB})$ and \linebreak $E_{RE}(\rho_{AB})$ respectively. They are
of a form typical of irreversible processes in that restoring the information
in $M$ is not sufficient to restore the original correlations between $M$ and
$AB$. In particular, they express that the loss of entanglement between Alice
and Bob at each stage must be accompanied by an even greater reduction in
mutual information between the memory and subsystems $AB$. This raises the
interesting open question of whether the inequalities (\ref{eq:ptom}) and
(\ref{eq:upper}) may be generalised to a relation of the kind%
\begin{equation}
E_{RE}(\rho_{A:(BM)})\leq E_{RE}(\chi_{A:(BM)})+I_{Q}(\rho_{M:(AB)}%
)-I_{Q}(\chi_{M:(AB)}) \label{eq:conj}%
\end{equation}
for any two entangled states, $\rho$ and $\chi$, where $\chi$ is obtained from
$\rho$ by any operation on the memory. This would give the physically
reasonable property that loss of the information in $M$ about $AB$ is always
greater than the loss of entanglement between the separated subsystems.

In summary, there are numerous decompositions of any bipartite mixed state
into a set of states $\rho_{i}$ with probability $p_{i}$. The average
entanglement of states in each decomposition is given by the relative entropy
of entanglement of the system extended by a memory whose orthogonal states are
classically correlated to the states of the decomposition. This correlation
records which state $\rho_{i}$ any member of an ensemble of mixed states
$\rho_{AB}^{\otimes n}$ is in. It is available to parties involved in
formation of the mixed state, but is not accessible to parties carrying out
distillation. When the classical information is fully available, different
decompositions give rise to different amounts of distillable entanglement, the
highest being entanglement of assistance and the lowest, entanglement of
formation. When access to the classical record is reduced, the amount of
distillable entanglement is reduced. In the limit where no information is
available, the distillable entanglement is given by the relative entropy of
entanglement of the state $\rho_{AB}$ itself, without the extension of the
classical memory. Our work shows that relative entropy of entanglement
provides a unifying measure for all cases, elucidating the role of classical
information and the appearance of irreversibility in manipulations of mixed
state entanglement.

\bigskip
\bibliographystyle{prsty}
\bibliography{two}

\begin{thebibliography}{10}

\bibitem{Bennett96}
C.~H. Bennett, H.~J. Bernstein, S. Popescu, and B. Schumacher, Phys. Rev. A
  {\bf 53},  2046  (1996).

\bibitem{Popescu97}
S. Popescu and D. Rohrlich, Phys. Rev. A {\bf 56},  R3319  (1997).

\bibitem{Vedral97PRL}
V. Vedral, M.~B. Plenio, M.~A. Rippin, and P.~L. Knight, Phys. Rev. Lett. {\bf
  78},  2275  (1997).

\bibitem{Vedral98}
V. Vedral and M.~B. Plenio, Phys. Rev. A {\bf 57},  1619  (1998).

\bibitem{Smolin96}
C.~H. Bennett, D.~P. DiVincenzo, J.~A. Smolin, and W.~K. Wootters, Phys. Rev. A
  {\bf 54},  3824  (1996).

\bibitem{Wootters98}
W.~K. Wootters, Phys. Rev. Lett. {\bf 80},  2245  (1998).

\bibitem{Hughston93}
L.~P. Hughston, R. Jozsa, and W.~K. Wootters, Phys. Lett. A {\bf 183},  14
  (1993).

\bibitem{Schumacher95}
B. Schumacher, Phys. Rev. A {\bf 51},  2738  (1995).

\bibitem{Lo99}
H. Lo and S. Popescu, quant-ph/9902045  (1999).

\bibitem{Lindblad75}
G. Lindblad, Comm. Math. Phys. {\bf 40},  147  (1975).

\bibitem{Eisert99}
J. Eisert {\it et~al.}, quant-ph/ 9907021  (1999).

\bibitem{Garisto98}
R. Garisto and L. Hardy, quant-ph/9808007  (1998).

\bibitem{Divincenzo98}
D.~P. DiVincenzo {\it et~al.}, quant-ph/9803033  (1998).

\end{thebibliography}
\end{document}